# Block Copolymer Derived Multifunctional Gyroidal Monoliths for 3-D Electrical Energy Storage Applications

Jörg G. Werner[1,2]†, Gabriel G. Rodríguez-Calero[2], Héctor D. Abruña[2]\*, Ulrich Wiesner[1]\*

[1] Department of Materials Science and Engineering, Cornell University, Ithaca NY, USA.

[2] Department of Chemistry and Chemical Biology, Cornell University, Ithaca NY, USA.

\*Correspondence to: hda1@cornell.edu, ubw1@cornell.edu.

†Current addresses: School of Engineering and Applied Sciences, Harvard University, Cambridge MA, USA.

**Multifunctional three-dimensional (3-D) nano-architectures, integrating all device components within tens of nanometers, offer great promise for next generation electrical energy storage applications, but have remained challenging to achieve. The lack of appropriate synthesis methods, enabling precise 3-D spatial control at the nanoscale, remains a key issue holding back the development of such intricate architectures. Here we present an approach to such systems based on the bottom-up synthesis of penta-continuous nanohybrid monoliths with four functional components integrated in a triblock terpolymer derived core-shell double gyroid architecture. Two distinct 3-D interpenetrating networks serving as cathode and current collector are separated from a carbon anode matrix by continuous, ultrathin polymer electrolyte shells. All periodically ordered domains are less than 20 nm in their layer dimensions and integrated throughout the macroscopic monolith. Initial electrochemical measurements with the Li-ion/S system exhibit reversible battery-like charge-discharge characteristics with orders of magnitude decreases in footprint area over conventional flat thin layer designs.**

Nanostructured materials have dramatically impacted fundamental and applied research as well as applications due to their unique properties arising from well-defined spatial confinement and large surface area-to-volume ratios[1–3]. Co-continuous nanohybrids of multiple distinct functional materials offer the potential for great advances in catalysis, energy conversion, and optical devices due to their large interfacial areas combined with three-dimensional (3-D) continuity of all phases[4–6]. While developments in top-down photolithographic techniques and materials have enabled access to ever decreasing feature sizes in two- and three-dimensional (2-D and 3-D) architectures for transistors and other circuit elements[7,8], leading to widespread deployment of electronics[9], rapid 3-D nanohybrid device formation remains challenging and often suffers from ill-defined phase dimensions and heterogeneous material distributions. The integration of multiple materials in a macroscopic 3-D nanostructure, with well-defined spatial arrangements and resolution below 50 nm through bottom-up chemical synthesis pathways, is expected to be more cost-effective than top-down methods. In particular, self-assembly represents a promising approach to overcome existing limitations[10,11]. Block copolymers (BCP) can self-assemble into 0-, 1-, 2-, and 3-D morphologies with tunable characteristic feature sizes at the nanoscale (5-50 nm)[12]. The cubic gyroid morphology, exhibiting multiple continuous phases, is a particularly promising structure with tremendous potential applications[13]. It is based on Schoen's G minimal surface that separates space into two interpenetrating 3-D continuous volumes with opposite chirality, which are related by inversion (Fig. 1a)[14]. In triblock terpolymer-derived core-shell double gyroids ($G^D$), this surface splits laterally, defining a third continuous matrix volume (black domain in Fig. 1b). This matrix separates two networks made from triple nodes connected by struts that have a core-shell structure (red and blue domains in Fig. 1b)[12]. The matrix plus two co-continuous core-shell network domains make the entire structure penta-continuous.

The concept of such a nanostructured 3-D solid-state architecture, with multiple distinct functional and interpenetrating components that comprise an entire electrical energy storage device, is especially attractive due to the expected increases in power and energy densities combined with enhanced safety[15]. Conceptually, such a device utilizes stacked nanometer-thin sheets of anode, electrolyte, cathode and current collectors enabling short ion diffusion distances. A 3-D device can be thought of as having these thin sheets folded up into the third dimension providing higher spatial efficiency than flat sheets, which require large footprint areas to contain an equivalent amount of material and stored energy. With the opposing electrodes only tens of nanometers apart, ion diffusion times decrease significantly (proportional to the square of diffusion length) potentially enabling high power output solely due to structural considerations and independent of materials properties[15]. Only the utilization of rigid 3-D nanostructures enables this close proximity of anode and cathode in a spatially efficient manner, since rolling of hypothetical nanometer thin layered sheets results in very high curvatures and would most likely yield in materials failure and short circuits between the electrodes. However, the lack of appropriate synthesis methods has held back the development of such devices. There are but a few reports addressing the assembly of all components of an electrical energy storage device into a single, 3-D integrated nanostructured architecture[16,17]. Successfully operating 3-D nanostructured energy storage architectures have only been demonstrated with electrode distances of microns, however, and the use of liquid electrolytes for ion-shuttling between them[18–20]. In particular, a proof-of-concept demonstration of the viability and function of such a device with all component dimensions below 50 nm remains elusive. Synthesis of such a multifunctional, co-continuous and interpenetrating energy storage architecture requires precise spatial control in synthesis and deposition of every individual functional component, which has proven extremely challenging. Furthermore, compatibility of all materials, synthesis chemistries and conditions is required for the successful integration of multiple functional materials with precise nanoscale control. In essence, two independently contacted, interpenetrating, electronically conducting redox-active nano-scale electrodes need to be homogeneously separated by a nanometer thick pin-hole free, electronically insulating but ionically conducting solid electrolyte layer. Self-assembly based periodically ordered 3-D BCP morphologies, such as the periodically ordered gyroidal networks that allow for precise control of pore size and materials' dimensions in the form of macroscopic monoliths, represent an enabling starting platform for such structures[21], due to homogeneous pore connectivity and size. Bottlenecks, typically present in disordered porous networks, would be detrimental to conformal coating and backfilling at this length scale. Such a bottom-up synthetic approach is in stark contrast to conventional composite electrical energy storage device fabrication, where electrodes, electrolyte/separator, and current collectors are prepared separately and then physically layered together, relaxing many of these demanding requirements, but sacrificing the clear advantages that accrue from a true 3-D structured device.

Here we demonstrate a synthesis concept that enabled the integration and interpenetration of four functional materials, at the nanoscale, with the above-mentioned properties into a penta-continuous and periodically ordered 3-D gyroidal device architecture. Our synthesis approach is based on block copolymer self-assembly directed double gyroidal mesoporous carbon ($G^DMC$) monolith formation[21]. $G^DMC$ monoliths simultaneously act as the anode active material, current collector, and structural framework for the subsequent surface restricted electrolyte polymerization and nano-confined cathode/current collector composite synthesis (Fig. 1). Homogeneity, mechanical properties, and transport characteristics make such gyroidal monoliths promising candidates for the exploration of fully nano-integrated 3-D electrical energy storage devices[22,23].

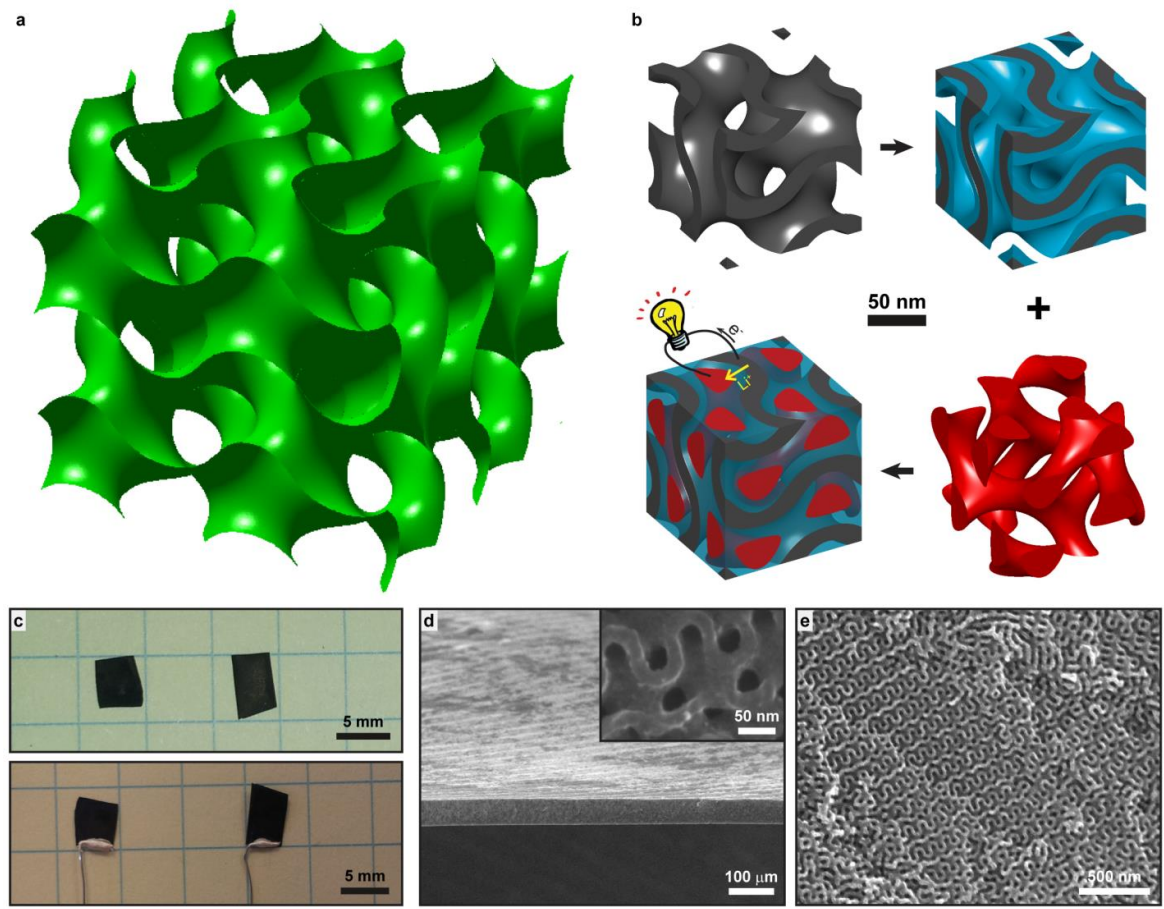

**Figure 1. Assembly of a penta-continuous interpenetrating and nanostructured hybrid from double gyroidal mesoporous carbon ($G^DMC$) monoliths. a**, Illustration of infinitely thin gyroid (G) minimal surface (green). **b**, Schematic illustration of the synthesis pathway: Redox-active $G^DMC$ electrode with the same structure as in **a**, but finite wall thickness (black; matrix) is conformally coated on both sides with nanoscaled polymer electrolyte (blue), here poly(phenylene oxide). Remaining two interpenetrating network mesopore channels are back-filled with composite electrodes (red) containing redox-active material and current collector, here sulfur and poly(3,4-ethylenedioxythiophene), respectively. **c**, Photographs of as-made $G^DMC$ monoliths (top) contacted in edge-on geometry (bottom). **d,e**, SEM images of a $G^DMC$ monolith exhibiting uniform thickness (**d**), surfaces with open and accessible gyroidal mesoporosity (**d**, inset), and uniform gyroidal cross-section (**e**).

It is important to emphasize that at this early stage of research into the synthesis of such 3-D interpenetrating nano-architectures for electrical energy storage, it was not realistic to expect the resulting monolithic devices to exhibit optimal electrochemical performance. Indeed, to the best of our knowledge, no reports exist to date of the synthesis of such complex penta-continuous and tetrafunctional 3-D gyroidal device nano-architectures described herein. Instead, our intent was to establish a "proof-of-principle" for overcoming a number of the existing critical hurdles. For example, with only nm pores remaining in a macroscopic 3-D anode scaffold, could chemical pathways be found to generate a conformal 10 nm thick separator coating throughout its mesoporous network? Could the resulting, even smaller, pores subsequently be backfilled with a composite that simultaneously functions as a cathode and current collector and remains physically separated from the initial anodic framework? With a separator reduced/limited to dimensions of only 10 nm, i.e. orders of magnitude thinner than those in conventional batteries, would it be possible to observe a stable potential across the fully assembled tetrafunctional monolithic device? Finally, could even a couple of charge/discharge cycles be run for such devices in which lithium ions travel only 10s of nanometers while electrons travel over macroscopic dimensions through an external electrical circuit to provide electrical energy? In this report, we demonstrate that such complex penta-continuous, tretrafunctional and interpenetrating gyroidal solid-state hybrid architectures can be successfully synthesized. Furthermore, we show how such monolithic device architectures display battery-like redox behavior and charge/discharge profiles with stable open circuit voltage, thereby establishing proof-of-principle. Despite non-optimal behavior, including lower than expected capacity of the gyroidal 3-D nanostructured device, measured capacities were still almost three orders of magnitude higher when compared to a theoretical flat (2-D) architecture with the same nanoscale dimensions and footprint area.

**Results and Discussion**

**Gyroidal mesoporous carbon monoliths as simultaneous anode active material and current collector.** Conducting $G^DMC$ materials were synthesized through self-assembly of a large molar mass triblock terpolymer (130 kDa, Supplementary Table 1) poly(isoprene)-*block*-poly(styrene)-*block*-poly(ethylene oxide) (PI-*b*-PS-*b*-PEO; ISO) in the presence of oligomeric phenol-formaldehyde resols as carbon precursor. Pyrolysis of the polymeric-organic hybrid led to degradation of the block copolymer and carbonization of the resols, resulting in an ordered 3-D mesoporous carbon network with large, homogeneous mesopores 40 nm in diameter, large porosity of 63 vol%, and carbon wall thickness of 15 nm (Figs. 1c-e, Supplementary Fig. 1, Supplementary Table 2). Our procedure yielded free-standing, monolithic $G^DMCs$ with a homogeneous thickness chosen to be approximately 70 microns and accessible mesoporosity from all surfaces (Fig. 1d). Comprehensive synthesis and characterization details for such carbons are described elsewhere[21]. The geometry led to an average carbon area density of 5 mg cm$^{-2}$, although the thickness is tunable to obtain a desired areal carbon loading[21]. Free-standing G$^D$MC monoliths were wire-contacted electronically in an edge-on geometry (Fig. 1c), rendering all surfaces electrochemically accessible for the subsequent surface-confined polymer electrolyte synthesis, and serving as the anode contact of the final device.

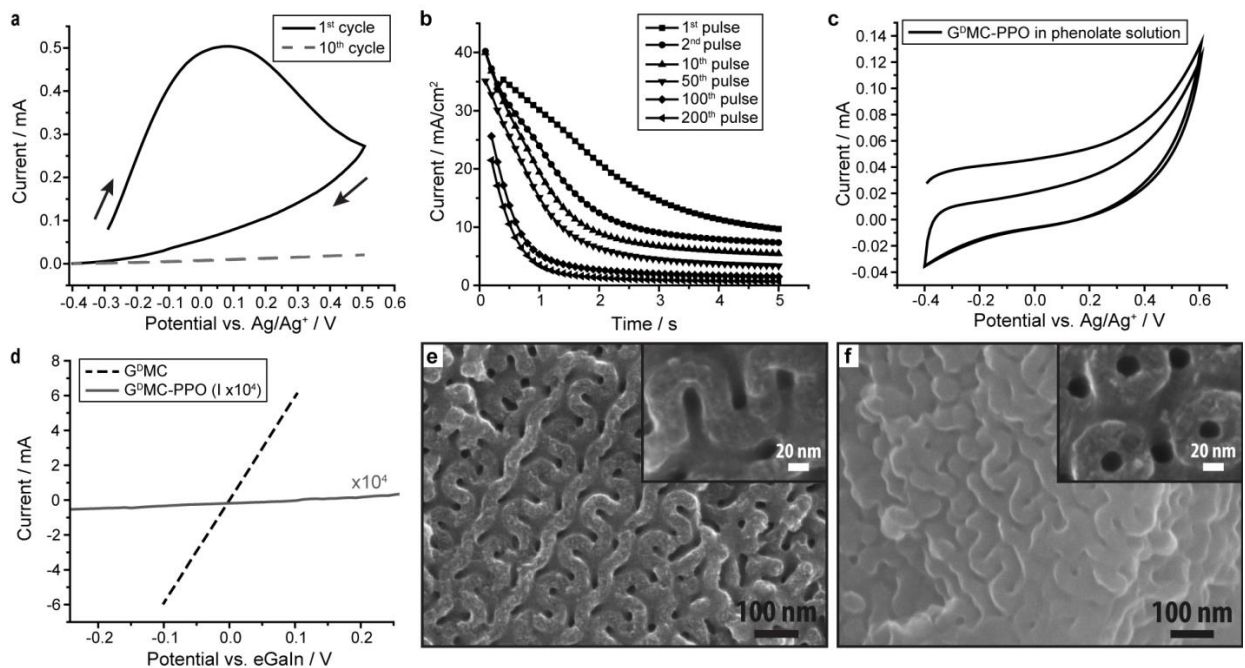

**Figure 2. Electropolymerization of ultrathin solid polymer electrolyte on monolithic mesoporous G$^D$MC electrode. a**, Cyclic voltammogram (CV) of G$^D$MC electrode in the poly(phenylene oxide) (PPO) electropolymerization solution. The broad oxidation peak of the first cycle, associated with PPO electropolymerization, disappears completely after 10 cycles. Arrows indicate potential sweep direction. **b**, Current-time traces of selected potentiostatic PPO electrodeposition pulses at +0.6 V vs. Ag/Ag$^+$ showing the decrease in oxidation current. **c**, CVs of PPO-coated G$^D$MC electrodes in phenolate solution after pulsed electropolymerization exhibiting only double layer characteristics. **d**, Two-electrode linear voltage sweep resistance measurements of the wire-connected G$^D$MC before (dashed, black) and after (solid, grey) PPO electropolymerization exhibiting an increase in resistance of more than 5 orders of magnitude. Note that the current of the PPO coated G$^D$MC was multiplied by a factor of 10$^4$ to facilitate comparison. **e**,**f**, SEM images of G$^D$MC anode after pulsed potentiostatic electropolymerization of PPO. Freshly cleaved cross-sections are depicted in (**e**), the monolith surface in (**f**), and high-magnification images in the insets.

**Conformal ultrathin solid electrolyte/separator coating.** G$^D$MC monoliths were conformally coated (Fig. 1b top) with an ultra-thin solid polymer electrolyte layer, poly(phenylene oxide) (PPO), of less than 10 nm in thickness using electropolymerization of phenol yielding an insulating PPO film and self-limiting film growth (Fig. 2a). Due to the propensity of deposition on previously uncoated areas (the current density on uncoated areas is much larger than on coated ones), self-regulating conformal growth was obtained until the entire accessible surface was homogeneously covered[24]. PPO had been previously shown to have the capability for forming a pinhole-free, ultrathin ion-conducting but electronically insulating layer, with selective molecular permeability tunable through the synthesis conditions[25]. In order to afford thin, dense, and homogenous PPO films on the two continuous gyroidal carbon surfaces, we used pulsed potentiostatic electropolymerization at +0.6 V vs. Ag/Ag$^+$, during which the oxidation current decayed substantially, while the double layer current remained at the application of the pulse-potential with attenuation (Fig. 2b). Subsequent cyclic voltammograms of PPO coated G$^D$MCs in phenol solution showed no oxidation peak, demonstrating the electrochemical insulation of the entire interior and exterior carbon surfaces (Fig. 2c). The resulting tricontinuous and bifunctional (anode + electrolyte/separator) monoliths were extensively characterized using scanning electron microscopy (SEM), confirming deposition of thin films on the G$^D$MCs (Figs. 2e,f, Supplementary Fig. 2d). The thickness of the PPO films was around 8-10 nm, estimated from the wall-to-wall distance (SEM) before and after electropolymerization, with remaining mesopores of 16-18 nm in size. The insulating, ultra-thin, PPO polymer electrolyte layers were further evaluated using two-probe resistance and lithiation measurements. An increase in resistance of more than 5 orders of magnitude, following PPO deposition, was measured via liquid metal surface contacts (Fig. 2d). Reversible lithiation/delithiation of PPO coated G$^D$MCs was achieved in two-electrode liquid cells against lithium metal, demonstrating the known permeability of the PPO layer to lithium ions in our gyroidal architecture (Supplementary Fig. 3c). The observed reversible charge capacity of 1 mAh cm$^{-2}$ translated to a specific capacity of 200 mAh g$^{-1}$ of the monolithic gyroidal PPO-coated carbon anode. This was in good agreement with the reversible capacity obtained from the G$^D$MC material without PPO coating in a standard coin cell test (Supplementary Fig. 3a,b). Critical for further synthesis steps, monolith surfaces remained open after PPO polymerization, maintaining accessibility of the inner mesoporosity (Fig. 2f).

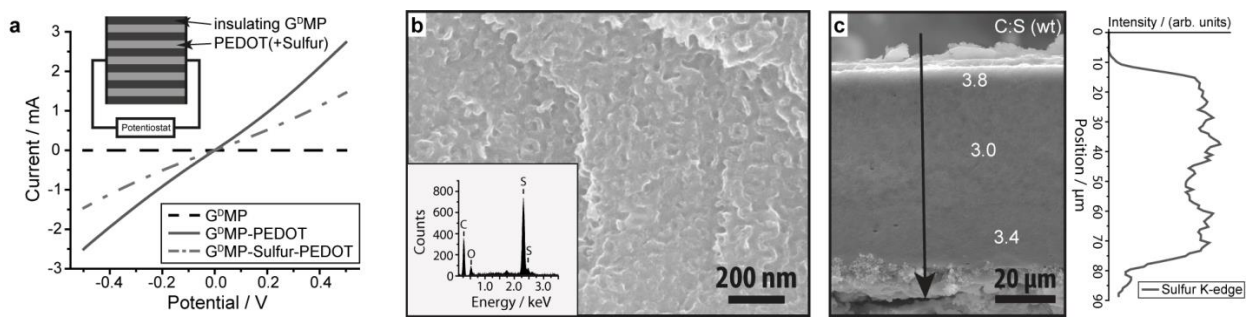

**Figure 3. Cathode composite infiltration. a**, Two-electrode resistance measurements of insulating double gyroidal mesoporous polymer monoliths (G$^D$MP, dashed black), back-filled with PEDOT (solid grey), and sulfur-PEDOT composite (dash-dotted grey). Inset: Schematic representation of resistance measurement. **b**,**c**, Cross-sectional SEM images of a sulfur-PEDOT backfilled PPO-coated G$^D$MC at high (**b**) and low (**c**) magnification with corresponding EDS spectrum (**b**) and EDS line scan (**c**) of the sulfur K-edge signal across the film (line scan position indicated by arrow in SEM and shown in Supplementary Fig. 4b). Numbers in (**c**) are the corresponding EDS-derived carbon:sulfur weight ratios at the respective positions.

**Infiltration with bifunctional composite as cathode active material and current collector.** PPO-coated bifunctional G$^D$MC monoliths were infiltrated with a bifunctional composite, completing the tetrafunctional and penta-continuous interpenetrating nanostructure (Fig. 2b bottom). Sulfur was chosen as a redox-active functional material. It exhibits a discharge voltage of 2-2.5 V versus lithium and is efficiently infiltrated into pores down to the microporous regime (<2 nm) using liquid/vapor infiltration at moderate temperatures (155 °C)[26]. This made sulfur a preferred cathode material for our solid-state interpenetrating 3-D gyroidal hybrids over commonly employed oxides such as lithium cobalt oxide (LCO). Synthesis conditions of LCO necessitate high temperatures to obtain the required crystal structure[27]. The presence of thin polymer electrolyte (PPO) layers excluded these or other aggressive conditions. Furthermore, external synthesis of dispersible lithium metal oxide nanoparticles small enough (<5nm) for effective infiltration into the 15-20 nm mesopore networks remains challenging. Finally, lithium metal oxides exhibit high lithiation potentials (>3.5 V vs. Li/Li$^+$), exceeding the limit of our polymer electrolyte[24]. While sulfur fulfilled all requirements as a cathode material, it is electronically insulating. We therefore chose to use the electronically conducting polymer poly(3,4-ethylenedioxythiophene) (PEDOT) as an integrated nanostructured current collector. We utilized an *in-situ* infiltration-polymerization method, since conducting polymer dispersion particles were too large for efficient mesopore infiltration. To this end, the monomer, EDOT, and the oxidizing reagent, iron(III) *para*-toluenesulfonate, were infiltrated as small molecular entities in solution that upon drying chemically polymerized EDOT to PEDOT inside the PPO-coated and sulfur infiltrated mesopores of the G$^D$MCs. In contrast to the anode side, the cathode side was thus constructed from two distinct materials acting as redox-active component (sulfur) and current collector (PEDOT), respectively, substantially increasing demands on chemical pathways and final device complexity relative, e.g. to trifunctional interpenetrating gyroidal solar cell devices demonstrated earlier[23]. While we did not anticipate homogeneous filling/infiltration of the entire monoliths with the sulfur-PEDOT composite, allowing quantitative access to all charges everywhere in the macroscopic devices, we hoped that it would be sufficient to demonstrate proof-of-principle performance characteristics of such tetrafunctional and penta-continuous interpenetrating 3-D nano-assembled electrical energy storage devices.

We separately tested the conductivity of the 3-D integrated nanostructured PEDOT current collector prepared by this method. To that end, we independently synthesized insulating double gyroidal mesoporous polymer (G$^D$MP) monoliths by limiting the heat treatment of self-assembled ISO-resols hybrids to 450 °C (Supplementary Information). These insulating G$^D$MP monoliths were used to test percolation and conductivity of PEDOT chemically synthesized within the nanoscopic confinement of the gyroidal mesopores. Resistance across G$^D$MP monoliths measured before and after PEDOT polymerization decreased substantially (Fig. 3a), suggesting successful percolation of conductive PEDOT through the macroscopic insulating polymer framework.

Comprehensive SEM and energy dispersive X-ray spectroscopy (EDS) analysis suggested that infiltration of sulfur into the mesopores of PPO coated G$^D$MC monoliths (Figs. 3b,c and Supplementary Figs. 4b-d) was fairly homogeneous. With a carbon to sulfur weight ratio of three, and assuming the PPO shell volume to be equal to the carbon matrix volume with a PPO density of 1.06 g cm$^{-3}$, the estimated maximum areal sulfur loading was 2.4 mg cm$^{-2}$. This sulfur loading yielded a theoretical areal capacity of 4 mAh cm$^{-2}$ (1675 mAh g$^{-1}$ of sulfur), with around one third of that for just the first sulfur discharge plateau, matching the carbon capacity of 1 mAh cm$^{-2}$.

While surfaces still exhibited gyroidal features after sulfur infiltration (Supplementary Fig. 4a), indicating minimal levels of external sulfur, PEDOT overlayers of a few microns on either surface of the monolithic films could be readily distinguished after PEDOT infiltration (Fig. 3c). These were beneficial for subsequent contacting of the cathode composite phase.

To introduce lithium, the sulfur-PEDOT phase was electrochemically reduced using a liquid electrolyte in a two-electrode configuration (Fig. 4a,b). The discharge curve exhibited two plateaus around 2.4 V and 1.4 V vs. Li/Li$^+$, respectively (Fig. 4b). The first plateau at 2.4 V with a capacity of approximately 1 mAh cm$^{-2}$ matched well with the sulfur reduction potential to form long-chain polysulfides and is in agreement with the sulfur capacity estimated above, while the second plateau at 1.4 V was over 500 mV below the commonly observed one[26,28]. The large overpotential of the second plateau was anticipated, however, as PEDOT becomes insulating at these low potentials[29], adding ohmic resistance to the system. Additionally, the de-doping of PEDOT at low potentials can contribute to the low-potential capacity, explaining the higher than expected capacity for the second discharge plateau[29]. The "external" reduction-lithiation of the sulfur-PEDOT phase was followed by galvanostatically charging and discharging the two interpenetrating gyroidal electrode phases, versus each other, in the solid state.

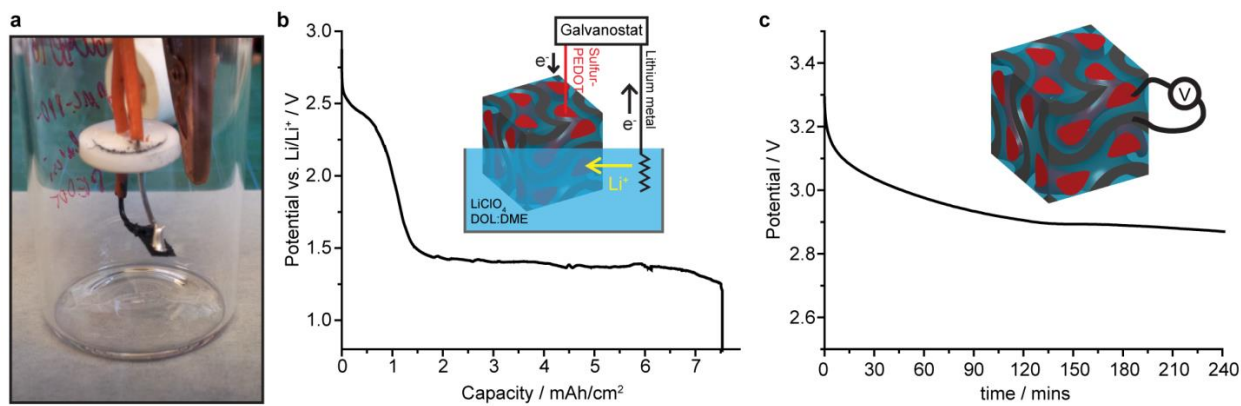

**Figure 4. Lithiation of penta-continuous tetrafunctional gyroidal nanostructured hybrids.** **a**, Photograph of a fully assembled tetrafunctional gyroidal nanohybrid device before external lithiation. **b**, Discharge/lithiation curve of the gyroidal sulfur-PEDOT phase vs. lithium metal in liquid electrolyte at 0.125 mA cm$^{-2}$. **c**, Open circuit voltage of the solid-state gyroidal hybrid over the first 4 hours after charge.

**Electrochemical properties.** The penta-continuous tetrafunctional gyroidal nanohybrid monoliths had to be charged to at least 3.5 V to exhibit a stable open-circuit voltage (Fig. 4c). The initial decrease of the open-circuit voltage over time after charging was assumed to originate from leakage currents above 3 V. However, most importantly is the demonstration of a stabilized open-circuit voltage after about 2 hours above 2.8 V over the 4 hours measured. Well-defined discharge plateaus were observed upon cycling the nanoscale 3-D interpenetrating anode and cathode phases (Fig. 5 and Supplementary Fig. 5). When discharged to 1 V at a current density of 0.125 mA cm$^{-2}$, the discharge curves exhibited a reversible plateau around 2.7 Volts with a capacity of up to 0.18 mAh cm$^{-2}$ and further irreversible capacity at lower potentials with a maximum capacity of 0.9 mAh cm$^{-2}$ (Fig. 5a). The reversibility of the first discharge plateau was confirmed by stable cycling up to ten times over a narrower voltage window above 1.5 V (Fig. 5b) with reversible capacities of around 0.2 mAh cm$^{-2}$. Cycling of the solid-state gyroidal nanohybrid battery at twice the rate (0.25 mAh cm$^{-2}$) for another 10 cycles, also yielded the reversible discharge plateau above 2.5 V with discharge capacities around 0.2 mA cm$^{-2}$, the same as at the lower rate, and irreversible discharge below 2.1 V (Fig. 5d).

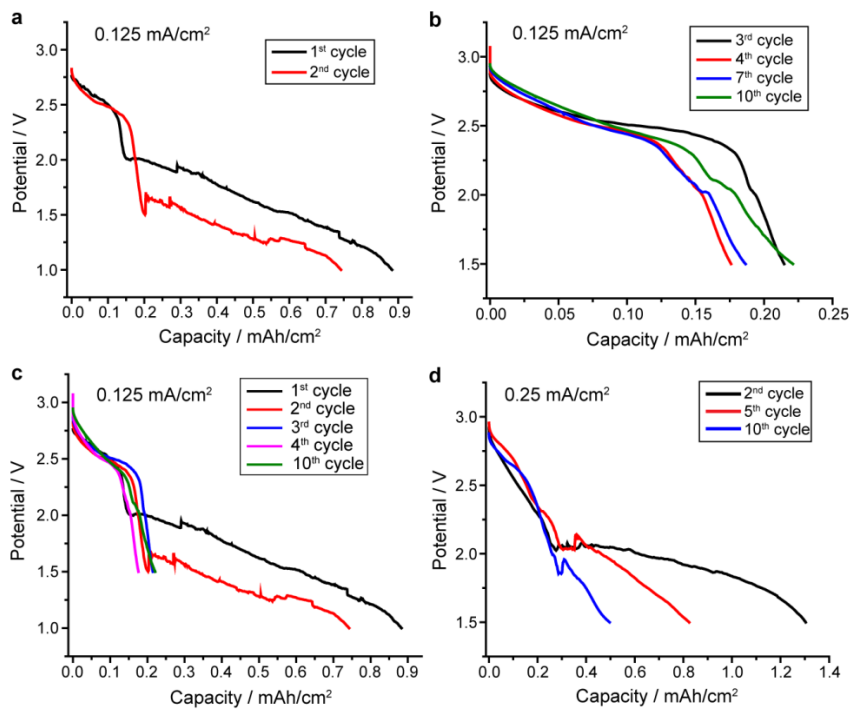

**Figure 5. Electrochemical characterization of penta-continuous tetrafunctional gyroidal nanostructured hybrids. a-c**, Discharge curves of a solid-state gyroidal hybrid at a current of 0.125 mA cm$^{-2}$: **a**, First two cycles with the lower cut-off potential at 1 V. **b**, Selected subsequent cycles with the lower cut-off potential at 1.5 V. **c**, Overlay of selected cycles. **d**, Discharge curves of selected cycles of a solid-state gyroidal hybrid at a current of 0.25 mA cm$^{-2}$ with the lower cut-off potential at 1.5 V.

The sulfur loading in the present architecture as estimated above would allow for capacity matching of the carbon anode with this first discharge plateau of elemental sulfur that lies within the conductivity window of PEDOT and with an overall approximate theoretical capacity of 1 mAh cm$^{-2}$. This capacity matching and utilization of only the first sulfur discharge plateau has the additional advantage that the volume expansion of the cathode is small as compared to the full lithiation of sulfur. Since there is no appreciable amount of sulfur on the top or bottom surfaces of the monoliths (*vide supra*), the achieved reversible capacity of our gyroidal tetrafunctional nanohybrids of approximately 20% of their theoretical capacity clearly arises from accessing material throughout the 3-D monoliths. This relatively low fraction suggests the formation of electronically disconnected parts of the nano-integrated cathode phase from the external current collector during or after lithiation. It is worth noting that the glass-vial set-up employed here for housing the 3-D nanohybrid energy storage device (Fig. 4a) was not optimized in any way and likely led to air contamination and subsequent failure of the device during the storage period after the 20 cycles described here. Despite these inefficiencies, which can likely be overcome through improved materials and device housing designs, our results clearly established proof-of-concept and feasibility of a self-assembly derived nanoscale 3-D interpenetrating cathode and anode architecture for electrical energy storage. This is particularly noteworthy in light of the fact that a solid-state separator only 10 nm in thickness, covering a large absolute area of approximately 400 cm$^2$ throughout tortuous macroscopic monolithic devices, was able to maintain an open circuit voltage of over 2.8 V for hours and a well-defined discharge plateau for up to 20 cycles. Our results suggest that the separator, despite its small thickness, is sufficiently pinhole-free and impermeable to the sulfur-PEDOT cathode phase over this entire area to prevent short circuits. While at first this appears surprising, it is perhaps instructive to remember that biology uses self-assembly of lipids to maintain a potential difference of around 0.1 V across liquid-state cell membranes of about the same thickness throughout the entire body[30].

While these proof-of-principle results on 3-D gyroidal electrochemical devices are stimulating and promising, additional work is still required to carefully assess what improved materials choices and characteristics will lead to enhanced performance. In addition, the optimal solid-state separator thickness, required to prevent short circuits under long-time use conditions, needs to be identified. These are areas of current research in our laboratories.

**Conclusions**

We have demonstrated the bottom-up synthesis of a macroscopic 3-D solid-state electrochemical energy storage device architecture with precise spatial control over four distinct functional materials with individual domain dimensions of 20 nm or less. Our tetrafunctional and penta-continuous nanohybrids consist of two distinct, interpenetrating redox-active cathode electrode and current collector networks separated from a carbon anode matrix by an ultrathin, pinhole-free polymer electrolyte layer in a gyroidal core-shell architecture. The resulting devices, enabled by G$^D$MC monolith formation, allowed separate contacting of the carbon anode and sulfur cathode phases and maintained a stable open-circuit voltage and well-defined discharge plateau at 2.7 V. The 3-D architecture allows for a substantial decrease in footprint area, due to wrapping/integrating of all nanoscaled electrochemical components into interpenetrating networks, and exploiting the third dimension. Thus, it represents a truly atom-efficient approach. As an illustrative comparison, keeping layer dimensions and materials the same, a flat, three-layer (anode, separator, cathode) battery design, with comparable capacity, would take up an area that would be 4,700 times larger. Even at only 20% capacity utilization as demonstrated here, this would still account for a device about 940 times larger in area, almost 3 orders of magnitude. All components of our synthesis concept were chosen to be mutually compatible with facile laboratory processing. They are either polymer chemistry based (polymer self-assembly based anode formation; electropolymerization of solid electrolyte; solution polymerization of cathode current collector) or moderate temperature processable (active cathode material). In particular, the synthesis does not include any expensive or time consuming top-down nanofabrication processes, making the approach accessible to a wider community. We hope that this work will spur interest in further development and improvement of multifunctional, fully integrated 3-D nanohybrid architectures for electrical energy storage.

**Methods**

**Gyroidal mesoporous carbon (G$^D$MC).** The G$^D$MC monoliths were synthesized using block copolymer (BCP) self-assembly. The structure directing amphiphilic triblock terpolymer poly(isoprene)-*block*-poly(styrene)-*block*-poly(ethylene oxide) (PI-*b*-PS-*b*-PEO, ISO) was synthesized via a step-wise anionic polymerization[31]. The characteristics of the ISO are summarized in Supplementary Table 1. Phenol-formaldehyde resols were used as carbon precursors and synthesized using oligomerization of phenol and formaldehyde in a molar ratio of 1:2 under basic conditions[21]. Monolithic ISO:resols hybrids with double gyroidal ($G^D$) morphology were obtained by dissolution of both components in a 1:0.54 weight ratio in tetrahydrofuran and chloroform (1:1 by weight), followed by solvent evaporation and annealing. The polymer-organic hybrid monoliths were plasma treated to remove surface capping layers before carbonization under inert gas flow above 1000 °C. Free-standing $G^D$MC monoliths were cut into rectangular shapes with footprint areas of 0.08-0.14 cm$^2$, attached to a wire in an edge-on geometry with silver epoxy (EPO-TEK H20E from EMS), and the contact was sealed with silicone rubber (Momentive RTV 108). Insulating gyroidal mesoporous polymer ($G^D$MP) materials were obtained from the same ISO-resols hybrids as the $G^D$MCs, but were submitted to heat treatment at only 450 °C under nitrogen for 3 hours yielding insulating phenolic resins[32].

**Electrodeposition of polymer electrolyte.** The walls and surfaces of the conductive and redox-active $G^D$MC electrodes were conformally coated with poly(phenylene oxide) (PPO) using self-limiting electropolymerization. The polymerization solution in acetonitrile was 0.05 M in phenol and tetramethylammonium hydroxide pentahydrate, and 0.1 M in tetrabutylammonium perchlorate (TBAP). Electropolymerization was conducted in a three-electrode set up with a platinum counter electrode and a Ag/Ag$^+$ reference electrode (silver wire in acetonitrile with 0.05 M silver perchlorate and 0.1 M TBAP) under an inert atmosphere using a Metrohm Autolab PGSTAT204 potentiostat. Phenol was oxidatively polymerized using pulsed potentiostatic deposition with 200 pulses at +0.6 V vs. Ag/Ag$^+$ for a duration of 5 sec with a 10 sec equilibration between each pulse[25]. Pulsed polymer electrodeposition was followed by cyclic voltammetric sweeps between -0.4 V and +0.6 V vs. Ag/Ag$^+$ at 20 mV s$^{-1}$. The mesoporous PPO-coated carbon monoliths were subsequently cleaned with ethanol and dried at room temperature in air.

**Sulfur infiltration.** Sulfur, employed as the cathode material, was introduced into the mesopores of PPO-coated $G^D$MCs through liquid/vapor infiltration at 155 °C for 24 hours. To that end, excess sulfur powder (1-3 mg) was put on and around the monoliths (with footprint areas of 0.08-0.14 cm$^2$, see above) in a sealed glass container, and heated to 155 °C for 24 hours.

**Integrated polymeric current collector infiltration.** Poly(ethylenedioxythiophene) (PEDOT) was chemically polymerized inside the PPO-coated and sulfur infiltrated G$^D$MC monolith mesopores using oxidative polymerization. A 0.7 M iron (III) para-toluenesulfonate solution in ethanol was freshly prepared and EDOT was added to make a 1 M EDOT solution[33]. The G$^D$MC-PPO-sulfur monoliths were immersed in the solution for 20 mins at 4 °C. The 3-D nanohybrid was subsequently dried in air at room temperature and 80 °C. The sulfur-PEDOT nanostructured composite phase was then contacted with silver epoxy and sealed with silicone rubber.

**Sulfur lithiation and galvanostatic testing.** The contacted tetrafunctional monolithic nanohybrid was immersed horizontally (i.e. parallel to the gas-liquid interface) in 1 M lithium perchlorate in a mixture of dioxolane (DOL) and dimethoxyethane (DME) (1:1 by volume) together with lithium foil in a septum capped vial under an argon atmosphere. The sulfur-PEDOT phase was discharged to 1 V vs. Li/Li$^+$ at a current of 0.125 mA cm$^{-2}$. After lithiation of the sulfur-PEDOT nanocomposite, the lithium foil was disconnected, and the penta-continuous and tetrafunctional nanohybrids were cycled with the nanostructured G$^D$MC and sulfur-PEDOT phases connected as anode and cathode, respectively, with varying cut-off voltages as described in the text after removal of the liquid electrolyte.

**Characterization.** Scanning electron microscopy (SEM) of carbonized samples was carried out on a Zeiss LEO 1550 FE-SEM or a Tescan Mira SEM operating at an accelerating voltage of 10-20 kV. The SEM was equipped with a Bruker energy dispersive spectrometer (EDS) for elemental analysis. SAXS measurements were performed on monolithic parent hybrid and resulting carbon materials at the Cornell High Energy Synchrotron Source (CHESS). Nitrogen sorption isotherms were obtained on a Micromeritics ASAP 2020 surface area and porosity analyzer at -196 °C. Lithiation and galvanostatic tests of monolithic carbon materials and tetrafunctional nanohybrids were executed using a BST8-WA 8-channel battery analyzer from MTI Corporation. Resistance through the PPO thin film of the PPO-coated G$^D$MC monoliths and of the PEDOT and sulfur-PEDOT infiltrated insulating gyroidal mesoporous polymer (G$^D$MP) frameworks was measured using cyclic voltammetry at a scan rate of 50 mV s$^{-1}$ with an AUTOLAB Metrohm Autolab PGSTAT204. The second contact for the uncoated and PPO-coated carbon monoliths was made using a liquid gallium-indium eutectic contact on one of the surfaces. Contacts for the PEDOT and sulfur-PEDOT infiltrated insulating G$^D$MP framework were made with silver epoxy (EPO-TEK H20E from EMS) on both surfaces.

**References**

1. Alivisatos, A. P. Semiconductor Clusters, Nanocrystals, and Quantum Dots. *Science* **271,** 933–937 (1996).

2. Xia, Y. *et al.* One-Dimensional Nanostructures: Synthesis, Characterization, and Applications. *Adv. Mater.* **15,** 353–389 (2003).

3. Aricò, A. S., Bruce, P., Scrosati, B., Tarascon, J.-M. & van Schalkwijk, W. Nanostructured materials for advanced energy conversion and storage devices. *Nat. Mater.* **4,** 366–377 (2005).

4. Llordés, A., Garcia, G., Gazquez, J. & Milliron, D. J. Tunable near-infrared and visible-light transmittance in nanocrystal-in-glass composites. *Nature* **500,** 323–326 (2013).

5. Joo, S. H. *et al.* Ordered nanoporous arrays of carbon supporting high dispersions of platinum nanoparticles. *Nature* **412,** 169–172 (2001).

6. Yella, A. *et al.* Porphyrin-sensitized solar cells with cobalt (II/III)-based redox electrolyte exceed 12 percent efficiency. *Science* **334,** 629–634 (2011).

7. Chou, S. Y., Krauss, P. R. & Renstrom, P. J. Imprint Lithography with 25-Nanometer Resolution. *Science* **272,** 85–87 (1996).

8. Ito, T. & Okazaki, S. Pushing the limits of lithography. *Nature* **406,** 1027–1031 (2000).

9. Franklin, A. D. Nanomaterials in transistors: From high-performance to thin-film applications. *Science* **349,** aab2750 (2015).

10. Whitesides, G. M. & Grzybowski, B. Self-assembly at all scales. *Science* **295,** 2418–2421 (2002).

11. Orilall, M. C. & Wiesner, U. Block copolymer based composition and morphology control in nanostructured hybrid materials for energy conversion and storage: solar cells, batteries, and fuel cells. *Chem. Soc. Rev.* **40,** 520–535 (2011).

12. Bates, F. S. & Fredrickson, G. H. Block Copolymers—Designer Soft Materials. *Phys. Today* **52,** 32–38 (1999).

13. Hajduk, D. A. *et al.* The gyroid- a new equilibrium morphology in weakly segregated diblock copolymers. *Macromolecules* **27,** 4063–4075 (1994).

14. Schoen, A. Infinite periodic minimal surfaces without self-intersections. *Nasa Tech. Note* D-5541 (1970).

15. Rolison, D. R. *et al.* Multifunctional 3D nanoarchitectures for energy storage and conversion. *Chem. Soc. Rev.* **38,** 226–252 (2009).

16. Ergang, N. S., Fierke, M. A., Wang, Z., Smyrl, W. H. & Stein, A. Fabrication of a Fully Infiltrated Three-Dimensional Solid-State Interpenetrating Electrochemical Cell. *J. Electrochem. Soc.* **154,** A1135 (2007).

17. Rhodes, C. P., Long, J. W., Pettigrew, K. a, Stroud, R. M. & Rolison, D. R. Architectural integration of the components necessary for electrical energy storage on the nanoscale and in 3D. *Nanoscale* **3,** 1731–1740 (2011).

18. Pikul, J. H., Gang Zhang, H., Cho, J., Braun, P. V & King, W. P. High-power lithium ion microbatteries from interdigitated three-dimensional bicontinuous nanoporous electrodes. *Nat. Commun.* **4,** 1732 (2013).

19. Liu, C. *et al.* An all-in-one nanopore battery array. *Nat. Nanotechnol.* 1–9 (2014). doi:10.1038/nnano.2014.247

20. Sun, K. *et al.* 3D printing of interdigitated Li-ion microbattery architectures. *Adv. Mater.* **25,** 4539–4543 (2013).

21. Werner, J. G., Hoheisel, T. N. & Wiesner, U. Synthesis and characterization of gyroidal mesoporous carbons and carbon monoliths with tunable ultralarge pore size. *ACS Nano* **8,** 731–743 (2014).

22. Cho, B.-K., Jain, A., Gruner, S. M. & Wiesner, U. Mesophase Structure-Mechanical and Ionic Transport Correlations in Extended Amphiphilic Dendrons. *Science* **305,** 1598–1601 (2004).

23. Crossland, E. J. W. *et al.* A Bicontinuous Double Gyroid Hybrid Solar Cell. *Nano Lett.* **9,** 2807–2812 (2009).

24. Rhodes, C. P., Long, J. W., Doescher, M. S., Fontanella, J. J. & Rolison, D. R. Nanoscale Polymer Electrolytes : Ultrathin Electrodeposited Poly ( Phenylene Oxide ) with Solid-State Ionic Conductivity. *J. Phys. Chem. B* **108,** 13079–13087 (2004).

25. McCarley, R. L., Irene, E. a. & Murray, R. W. Permeant molecular sieving with electrochemically prepared 6-nm films of poly(phenylene oxide). *J. Phys. Chem.* **95,** 2492–2498 (1991).

26. Ji, X., Lee, K. T. & Nazar, L. F. A highly ordered nanostructured carbon–sulphur cathode for lithium–sulphur batteries. *Nat. Mater.* **8,** 500–506 (2009).

27. Li, X., Cheng, F., Guo, B. & Chen, J. Template-Synthesized $LiCoO_2$ , $LiMn_2O_4$ , and $LiNi_{0.8}Co_{0.2}O_2$ Nanotubes as the Cathode Materials of Lithium Ion Batteries. *J. Phys. Chem. B* **109,** 14017–14024 (2005).

28. Cuisinier, M. *et al.* Sulfur Speciation in Li–S Batteries Determined by Operando X-ray Absorption Spectroscopy. *J. Phys. Chem. Lett.* **4,** 3227–3232 (2013).

29. Burkhardt, S. E. *et al.* Theoretical and electrochemical analysis of poly(3,4-alkylenedioxythiophenes): Electron-donating effects and onset of p-doped conductivity. *J. Phys. Chem. C* **114,** 16776–16784 (2010).

30. Stein, W. D. & Lieb, W. R. *Transport and diffusion across cell membranes*. (Academic Press, Orlando, 1986).

31. Bailey, T. S., Hardy, C. M., Epps, T. H. & Bates, F. S. A Noncubic Triply Periodic Network Morphology in Poly(isoprene- *b* -styrene- *b* -ethylene oxide) Triblock Copolymers. *Macromolecules* **35,** 7007–7017 (2002).

32. Meng, Y. *et al.* A family of highly ordered mesoporous polymer resin and carbon structures from organic-organic self-assembly. *Chem. Mater.* **18,** 4447–4464 (2006).

33. Hong, K. H., Oh, K. W. & Kang, T. J. Preparation and properties of electrically conducting textiles by in situ polymerization of poly(3,4-ethylenedioxythiophene). *J. Appl. Polym. Sci.* **97,** 1326–1332 (2005).

**Acknowledgments.** We thank Dr. Kahyun Hur for assistance with the graphics and Dr. J. Gao for assistance with the coin cell test. This work was primarily supported as part of the Energy

Materials Center at Cornell (emc$^2$), an Energy Frontier Research Center funded by the U.S. Department of Energy, Office of Science, Basic Energy Sciences under Award no. DE-SC0001086. This work made use of the Cornell Center for Materials Research Shared Facilities supported through the NSF MRSEC program (DMR-1120296). This work was further based upon research conducted at the Cornell High Energy Synchrotron Source (CHESS) supported by the National Science Foundation and the National Institutes of Health/National Institute of General Medical Sciences under NSF award no. DMR-1332208. J.G.W. acknowledges partial support from the National Science Foundation (NSF) under the award no. DMR-1409105.

**Author Contributions.** All authors conceived the idea and concept for the project. J.G.W. and G.G.R.-C. performed electropolymerization experiments. J.G.W. prepared and characterized all other materials. J.G.W. performed electrochemical characterization of all individual components and of the solid-state 3-D gyroidal nanohybrids. J.G.W. and U.W. wrote the manuscript. All authors discussed the results and helped in improving the manuscript.

**Competing financial interests.** The authors declare no competing financial interests. A patent application related to this research has been filed by Cornell University.

**Materials & Correspondence.** Readers are welcome to comment on the online version of the paper. Correspondence and requests for materials should be addressed to U.W. (ubw1@cornell.edu) or H.D.A. (hda1@cornell.edu).

# Supplementary Information for

## Block Copolymer Derived Multifunctional Gyroidal Monoliths for 3-D Electrical Energy Storage Applications

Jörg G. Werner[1,2]†, Gabriel G. Rodríguez-Calero[2], Héctor D. Abruña[2]*, Ulrich Wiesner[1]*

*Correspondence to: hda1@cornell.edu, ubw1@cornell.edu

**The supplementary information includes:**

    Supplementary Figures 1 to 5
    Supplementary Tables 1 and 2
    Detailed experimental procedures

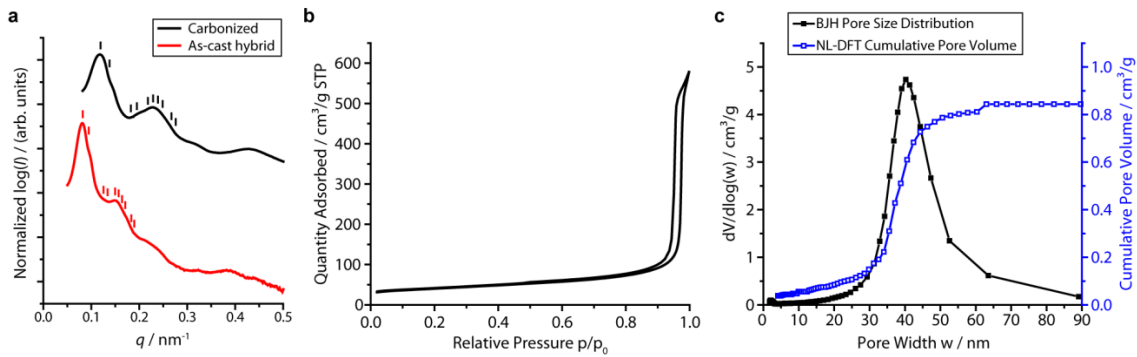

**Supplementary Figure 1. Gyroidal mesoporous carbon (G$^D$MC) characterization.** **a**, Small-angle X-ray scattering pattern of G$^D$MC as-cast hybrid (bottom, red) and after carbonization (black, top). The markings indicate expected positions for the *Ia$\bar{3}$d* space group lattice peaks of the double gyroid morphology with a (100) spacing of 188 nm (hybrid) and 129 nm (carbonized), respectively. **b**, Nitrogen sorption isotherms of carbonized G$^D$MC. **c**, BJH pore size distribution (solid symbols) and DFT cumulative pore volume (open, blue symbols) of carbonized G$^D$MC calculated from the nitrogen adsorption isotherm.

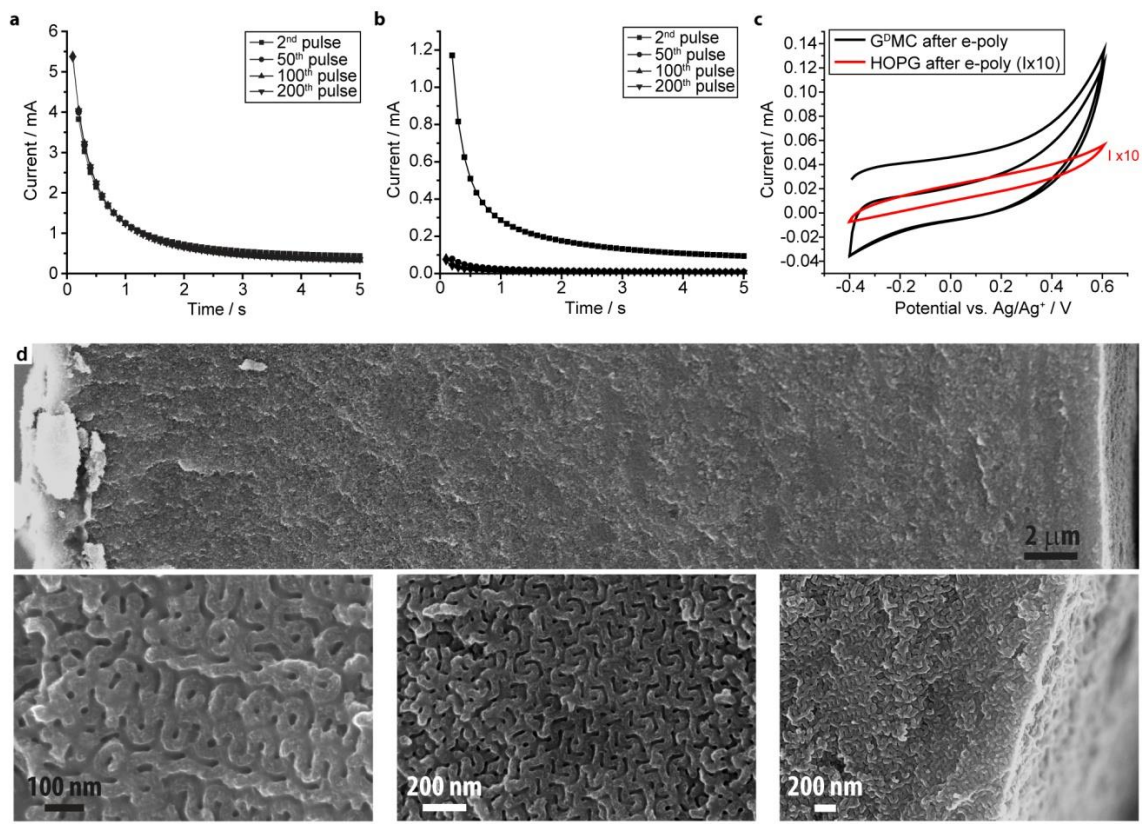

**Supplementary Figure 2. Electropolymerization of poly(phenylene oxide) (PPO).** **a**, Current-time traces of selected potentiostatic pulses at +0.6 V vs. Ag/Ag$^+$ of a G$^D$MC monolith in the supporting electrolyte demonstrating no change over 200 pulses. **b**, Current-time traces of selected potentiostatic deposition pulses of PPO on a flat carbon substrate (highly oriented pyrolytic graphite, HOPG) at +0.6 V vs. Ag/Ag$^+$ showing the very fast decrease in current with number of pulses. **c**, Cyclic voltammograms after pulsed potentiostatic PPO deposition in the same solutions of G$^D$MC and HOPG exhibiting no oxidation peak and much larger double layer current for G$^D$MC (current of the HOPG is multiplied by a factor of 10). **d**, Cross-sectional SEM images of PPO-coated G$^D$MC at low magnification (top) and at higher magnifications (bottom) close to the left surface, the film center, and the right surface (from left to right) suggesting homogeneous PPO coating of the gyroidal mesopores throughout the monolith.

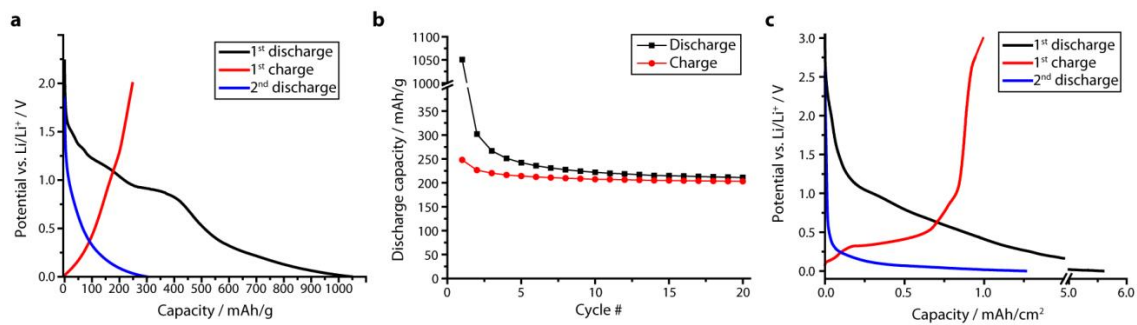

**Supplementary Figure 3. Galvanostatic measurements of $G^DMCs$. a**, Charge and discharge curves of powdered $G^DMC$ in a standard coin cell versus lithium metal at a current of 30 mA $g_C^{-1}$. **b**, Charge and discharge capacities of the first 22 cycles of the coin cell test demonstrating a reversible capacity of 220 mAh $g^{-1}$. **c**, Discharge and charge curves of PPO-coated $G^DMC$ at a current of 0.1 mA $cm^{-2}$ (corresponding to approximately 20 mA $g_C^{-1}$), demonstrating reversible lithiation-delithiation through the PPO-layer.

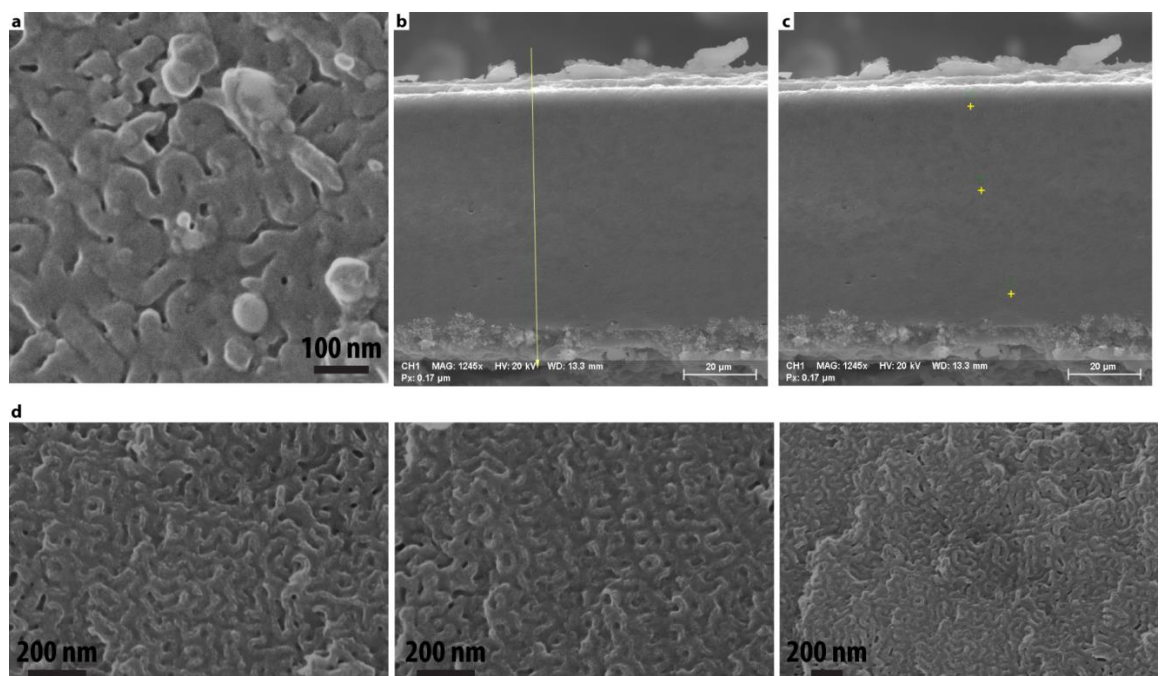

**Supplementary Figure 4. Tetrafunctional gyroidal 3-D nanohybrid characterization. a,** High-resolution SEM image of the PPO-coated G$^D$MC monolith surface after sulfur infiltration evidencing remaining porosity, which in turn signifies that no appreciable amount of excess external sulfur has deposited on the monolith surface during the sulfur infiltration process. **b,** Cross-sectional SEM image of the fully integrated tetrafunctional gyroidal 3-D nanohybrid described in the main text with the arrow showing the position of the line scan plotted in Fig. 3c. **c**, The same cross-sectional SEM image with the marks showing the position of quantified EDS-spectra with estimated carbon to sulfur weight ratios of 3.8, 3.0, and 3.4 at the top, middle, and bottom of the monolith, respectively. **d**, High-magnification SEM images of a penta-continuous and tetrafunctional gyroidal 3-D nanohybrid (G$^D$MC coated with PPO and back-filled with sulfur and PEDOT) at different positions across the monolith cross-section suggesting fairly homogeneous infiltration efficiency.

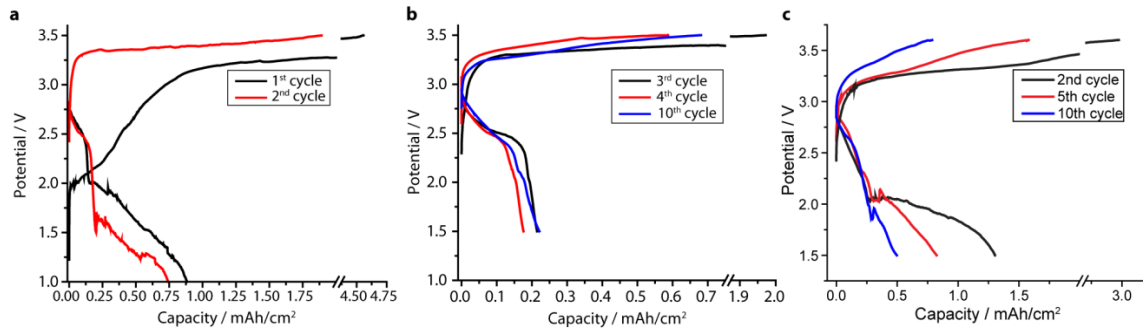

**Supplementary Figure 5. Galvanostatic characterization of penta-continuous tetrafunctional gyroidal energy storage nanohybrid monolith. a**, Solid-state charge-discharge curves of the first two cycles that were run over the potential window of 1-3.5 V at a current rate of 0.125 mA cm$^{-2}$. **b**, Charge and discharge curves of selected cycles that were run in the potential window of 1.5-3.5 V at a current rate of 0.125 mA cm$^{-2}$. **c**, Charge and discharge curves of selected cycles that were run in the potential window of 1.5-3.6 V at a current rate of 0.25 mA cm$^{-2}$.

**Supplementary Table 1.**

Triblock terpolymer ISO molar mass and composition.

| $M_n^*$ / kg/mol | $f_w^\dagger$ (PI) | $f_w^\dagger$ (PS) | $f_w^\dagger$ (PEO) | $M_w/M_n^\ddagger$ |
|---|---|---|---|---|
| 129.6 | 15.4% | 31.4% | 53.2% | 1.09 |

* Obtained from $^1$H-NMR and GPC analysis. $^\dagger$ Obtained from $^1$H-NMR. $^\ddagger$ Obtained from GPC.

**Supplementary Table 2.**

Properties of the gyroidal mesoporous carbon (G$^D$MC).

| Name | ISO:resols (weight ratio) | Mesopore surface area$^*$ | Pore volume$^\dagger$ | Porosity$^\ddagger$ | Unit cell size$^\S$ |
|---|---|---|---|---|---|
| G$^D$MC | 1 : 0.54 | 94 m$^2$/g | 0.86 cm$^3$/g | 63 vol% | 129 nm |

* Calculated from the difference of the BET surface area and the t-plot micropore surface area. $^\dagger$ Derived from the total adsorbed nitrogen volume at a relative pressure of 0.99. $^\ddagger$ Calculated using the pore volume and the specific volume of 0.5 cm$^3$/g for carbon. $^\S$ Derived from small angle X-ray scattering (SAXS).

**Detailed experimental procedures.**

**Gyroidal mesoporous carbon (G$^D$MC).** The G$^D$MC monoliths were synthesized using block copolymer (BCP) self-assembly. The structure directing amphiphilic triblock terpolymer poly(isoprene)-*block*-poly(styrene)-*block*-poly(ethylene oxide) (PI-*b*-PS-*b*-PEO, ISO) was synthesized via a step-wise anionic polymerization[1]. The characteristics of the ISO are summarized in Supplementary table 1. Phenol-formaldehyde resols were used as carbon precursors and synthesized using oligomerization of phenol and formaldehyde in a molar ratio of 1:2 under basic conditions[2]. Monolithic ISO:resols hybrids with double gyroidal (G$^D$) morphology were obtained by dissolution of both components in a 1:0.54 weight ratio in tetrahydrofuran (THF) and chloroform (1:1 by weight) and solvent evaporation at 50 °C in a Teflon dish. The dry films were annealed at 125 °C for 24 hours prior to carbonization. Irrespective of morphology, it is common in block copolymer (BCP) self-assembly to observe lamellar layers on surfaces of cast films due to surface energy differences between individual blocks and their corresponding affinity to air and the casting dish material[2]. These surface reconstituted lamellar capping layers were removed using an argon-oxygen plasma treatment for 40 mins on the polymeric-organic hybrid films, prior to carbonization. This treatment is essential to make the interior porosity accessible from the surfaces, after carbonization. Powdered G$^D$MC materials for characterization of the bulk properties, as well as G$^D$MC monoliths were carbonized at temperatures above 1000 °C under inert gas (argon/nitrogen) flow[2]. Characterization results and properties of the G$^D$MCs are shown in Fig. 1 and Supplementary Fig. 1, and summarized in Supplementary Table 2. The conductivity of the gyroidal mesoporous carbons ranged from 0.1 S cm$^{-1}$ to 2 S cm$^{-1}$ depending on the carbonization temperature[2]. Discussion of the detailed characterization of the conducting gyroidal mesoporous carbon materials can be found in Reference 2. Free-standing G$^D$MC monoliths were cut into rectangular shapes with footprint areas of 0.08-0.14 cm$^2$, attached to a wire in an edge-on geometry with silver epoxy (EPO-TEK H20E from EMS), and cured at 80 °C for at least 10 hours. The exposed silver contact was then sealed to the outside using silicone rubber adhesive sealant (Momentive RTV 108) and cured at room temperature for at least 24 hours. Insulating gyroidal mesoporous polymer (G$^D$MP) materials were obtained from the same ISO-resols hybrids as the G$^D$MCs, but were submitted to heat treatment at only 450 °C under nitrogen for 3 hours. At 450 °C, the phenol-formaldehyde resols form an insulating, cross-linked polymeric resin, while the structure directing ISO triblock terpolymer decomposes resulting in gyroidal mesoporous polymer materials[3].

**Electrodeposition of polymer electrolyte.** The walls and surfaces of the conductive and redox-active G$^D$MC electrodes were conformally coated with poly(phenylene oxide) (PPO) using self-limiting electropolymerization. The polymerization solution in acetonitrile was 0.05 M in phenol and tetramethylammonium hydroxide pentahydrate, and 0.1 M in tetrabutylammonium perchlorate (TBAP). The solution was kept under an inert gas atmosphere (nitrogen) to prevent oxidation of the deprotonated phenol. Wire-connected G$^D$MC monoliths were immersed completely into the solution and left soaking for 15 mins. Electropolymerization was conducted in a three-electrode set up with a platinum counter electrode and a Ag/Ag$^+$ reference electrode (silver wire in acetonitrile with 0.05 M silver perchlorate and 0.1 M TBAP) using a Metrohm Autolab PGSTAT204 potentiostat. Cyclic voltammetry (CV) of phenol in basic solution on G$^D$MCs showed a broad irreversible oxidation peak around 0.08 V vs. Ag/Ag$^+$ that decayed rapidly with continuous cycling (Fig. 2a). No indication of an oxidation peak nor large double layer current was present after 10 cycles, indicating blockage of the surface pores and mesoporosity due to the electropolymerization reaction. To afford a thin and dense PPO film, phenol was oxidatively polymerized using pulsed potentiostatic deposition (Fig. 2b). 200 pulses at +0.6 V vs. Ag/Ag$^+$, well above the peak oxidation current in the mass transport limit, were employed for a duration of 5 sec with a 10 sec equilibration between each pulse, to allow for sufficient monomer diffusion into the mesopores[4]. In comparison to the electropolymerization, the background response of the electrolyte showed only double layer current with very little (if any) changes in the current response over 200 pulses (Supplementary Fig. 2a). Pulsed polymer electrodeposition was followed by cyclic voltammetric sweeps between -0.4 V and +0.6 V vs. Ag/Ag$^+$ at 20 mV s$^{-1}$. These cyclic voltammograms (CVs) after potentiostatic PPO deposition in the polymerization solution showed no oxidation peak occurring at the oxidation potential of the phenolate ion (compare Fig. 2a and Fig. 2c). The double layer response of the PPO coated G$^D$MCs showed a much larger CV-area than a flat substrate with the same geometrical surface area (Supplementary Figs. 2b,c), corroborating retained accessibility of mesopores after PPO electropolymerization. The mesoporous PPO-coated carbon monoliths were subsequently soaked in ethanol to remove excess electrolyte and dried at room temperature.

**Lithiation-delithiation of $G^DMC$-electrolyte assembly.** PPO-coated $G^DMC$ monoliths and lithium foil were immersed in 1 M lithium perchlorate in dimethyl carbonate and ethylcarbonate (1:1 by volume) in a septum capped vial under an argon atmosphere. After soaking for 2 days, $G^DMC$s were galvanostatically lithiated/delithiated at 0.1 mA cm$^{-2}$ to 0 V and 2 V vs. Li/Li$^+$, respectively, in a two-electrode setup using an MTI BT8 battery tester. The first discharge (lithiation of carbon) curves exhibited a plateau around 1 V vs. Li/Li$^+$ and a very large capacity, as is commonly observed for the first discharge of carbon anode materials due to build-up of a solid-electrolyte interface (SEI, Supplementary Fig. 3)[5]. The following charge (delithiation) and discharge (lithiation) showed a capacity of 1 and 1.3 mAh cm$^{-2}$, respectively, demonstrating reversible lithium intercalation through the PPO electrolyte layer. Assuming a carbon loading of 5 mg cm$^{-2}$, the areal capacity corresponds to a specific capacity of approximately 200 mAh g$^{-1}$.

**Sulfur infiltration.** Sulfur, employed as the cathode material, was introduced into the mesopores of PPO-coated $G^DMC$s through liquid/vapor infiltration at 155 °C for 24 hours. To that end, excess sulfur powder (1-3 mg) was put on and around the monoliths (with footprint areas of 0.08-0.14 cm$^2$, see above) in a sealed glass container, and heated to 155 °C for 24 hours.

**Integrated polymeric current collector infiltration.** Poly(ethylenedioxythiophene) (PEDOT) was chemically polymerized inside the PPO-coated and sulfur infiltrated $G^DMC$ monolith mesopores using oxidative polymerization. A 0.7 M iron (III) para-toluenesulfonate solution in ethanol was freshly prepared and EDOT was added to make a 1 M EDOT solution[6]. The $G^DMC$-PPO-sulfur monoliths were immersed in the solution with the wire connection above the liquid-air interface and kept for 20 mins at 4 °C. The 3-D nanohybrid was subsequently removed from the solution, dried at room temperature for at least 4 hours, and further dried at 80 °C for at least 6 hours. The sulfur-PEDOT nanostructured composite phase was then contacted with silver epoxy (EPO-TEK H20E from EMS) on one of the monolith's surfaces. The cured electrical contact was again sealed to the outside with silicone rubber adhesive sealant (Momentive RTV 108) and cured for 2 days at room temperature.

**Sulfur lithiation and galvanostatic testing.** The contacted penta-continuous and tetrafunctional monolithic nanohybrid was immersed horizontally (i.e. parallel to the gas-liquid interface) in 1

M lithium perchlorate in a mixture of dioxolane (DOL) and dimethoxyethane (DME) (1:1 by volume) together with a lithium foil in a septum capped vial under argon atmosphere. The electrolyte was chosen to avoid unwanted side reactions with polysulfides that carbonate-based electrolytes undergo[7]. After soaking for 2 days, the sulfur-PEDOT phase was discharged to 1 V vs. Li/Li$^+$ at a current of 0.125 mA cm$^{-2}$ (Fig. 4a,b). After lithiation of the sulfur-PEDOT nanocomposite the lithium foil was disconnected, and the penta-continuous and tetrafunctional nanohybrids were cycled to 3 V at 0.125 mA cm$^{-2}$ with the nanostructured G$^D$MC and sulfur-PEDOT phases connected as anode and cathode, respectively. Subsequently, the electrolyte was removed and the tetrafunctional gyroidal nanohybrids were charged to 3.5 Volts and discharged at 0.125 mA cm$^{-2}$ with varying cut-off voltages as described in the text.

**Characterization.** Scanning electron microscopy (SEM) of carbonized samples was carried out on a Zeiss LEO 1550 FE-SEM or a Tescan Mira SEM operating at an accelerating voltage of 10-20 kV. The SEM was equipped with a Bruker energy dispersive spectrometer (EDS) for elemental analysis. SAXS measurements were performed on monolithic parent hybrid and resulting carbon materials at the Cornell High Energy Synchrotron Source (CHESS). The sample to detector distance was 2.6 m and the X-ray wavelength, $\lambda$, was 1.20 Å. The scattering vector, $q$, in the experiments was defined as $q = (4\cdot\pi/\lambda)\cdot\sin\theta$, where $\theta$ is half of the scattering angle. Nitrogen sorption isotherms were obtained on a Micromeritics ASAP 2020 surface area and porosity analyzer at -196 °C. Lithiation and galvanostatic tests of monolithic carbon materials and tetrafunctional nanohybrids were executed using a BST8-WA 8-channel battery analyzer from MTI Corporation. Resistances through the PPO thin film of the PPO-coated G$^D$MC monoliths and of the PEDOT and sulfur-PEDOT infiltrated insulating gyroidal mesoporous polymer (G$^D$MP) frameworks were measured using cyclic voltammetry at a scan rate of 50 mV s$^{-1}$ with an AUTOLAB Metrohm Autolab PGSTAT204. The second contact for the uncoated and PPO-coated carbon monoliths was made using a liquid gallium-indium eutectic contact on one of the surfaces. Contacts for the PEDOT and sulfur-PEDOT infiltrated insulating G$^D$MP framework were made with silver epoxy (EPO-TEK H20E from EMS) on both surfaces.

**References**
1. Bailey, T. S., Hardy, C. M., Epps, T. H. & Bates, F. S. A Noncubic Triply Periodic Network Morphology in Poly(isoprene-*b*-styrene-*b*-ethylene oxide) Triblock Copolymers. *Macromolecules* **35,** 7007–7017 (2002).
2. Werner, J. G., Hoheisel, T. N. & Wiesner, U. Synthesis and characterization of gyroidal mesoporous carbons and carbon monoliths with tunable ultralarge pore size. *ACS Nano* **8,** 731–743 (2014).
3. Meng, Y. *et al.* A family of highly ordered mesoporous polymer resin and carbon structures from organic-organic self-assembly. *Chem. Mater.* **18,** 4447–4464 (2006).
4. McCarley, R. L., Irene, E. a. & Murray, R. W. Permeant molecular sieving with electrochemically prepared 6-nm films of poly(phenylene oxide). *J. Phys. Chem.* **95,** 2492–2498 (1991).
5. Xing, W. *et al.* Synthesis of ordered nanoporous carbon and its application in Li-ion battery. *Electrochim. Acta* **51,** 4626–4633 (2006).
6. Hong, K. H., Oh, K. W. & Kang, T. J. Preparation and properties of electrically conducting textiles by in situ polymerization of poly(3,4-ethylenedioxythiophene). *J. Appl. Polym. Sci.* **97,** 1326–1332 (2005).
7. Gao, J., Lowe, M. A., Kiya, Y. & Abruña, H. D. Effects of Liquid Electrolytes on the Charge–Discharge Performance of Rechargeable Lithium/Sulfur Batteries: Electrochemical and in-Situ X-ray Absorption Spectroscopic Studies. *J. Phys. Chem. C* **115,** 25132–25137 (2011).